%% file: gm22HDM.tex
\documentclass[12pt]{article}
\pdfoutput=1
\usepackage[utf8]{inputenc} 
\usepackage{amsmath,epsfig,cite,latexsym,color,amssymb,etoolbox}
\usepackage{mathrsfs}
\usepackage{graphicx}
\usepackage{caption,subcaption}
\DeclareGraphicsExtensions{.pdf,.png,.jpg,.eps}
\usepackage{hyperref}

\newcommand{\thdm}{{\text{2HDM}}}
\newcommand{\amu}{a_\mu}
\newcommand{\amub}{a_{\mu}^{\text B}}
\newcommand{\amuf}{a_{\mu}^{\text F}}
\newcommand{\tb}{t_\beta}
\newcommand{\MZ}{M_{Z}}
\newcommand{\Mh}{h}
\newcommand{\MH}{H}
\newcommand{\MA}{A}
\newcommand{\MHpm}{H^{\pm}}

\newcommand{\MMh}{M_{\Mh}}
\newcommand{\MMH}{M_{\MH}}
\newcommand{\MMA}{M_{\MA}}
\newcommand{\MMHpm}{M_{\MHpm}}
\newcommand{\CBA}{c_{\beta\alpha}}
\newcommand{\SBA}{s_{\beta\alpha}}
\newcommand{\zl}{\zeta_l}
\newcommand{\zu}{\zeta_u}
\newcommand{\zd}{\zeta_d}
\newcommand{\az}{\amu^{\text{non-Yuk}}}

\begin{document}
\begin{center}
{\Large\bf\boldmath Muon $g-2$ in the 2HDM: maximum results \\[1ex] and
  detailed phenomenology}
\\\vspace{3em}
{Adriano Cherchiglia$^1$, Dominik St\"ockinger,$^2$ Hyejung St\"ockinger-Kim$^2$}\\[2em]
 {\sl $^1$ Universidade Federal do ABC - Centro de Ciências Naturais e
   Humanas, Santo Andr\'e - Brazil}
 \\
   {\sl $^2$Institut f\"ur Kern- und Teilchenphysik, TU Dresden, 01069 Dresden, Germany}
\end{center}
\vspace{2ex}
\begin{abstract}
We present a comprehensive analysis of the muon magnetic moment $\amu$
in the flavour-aligned two-Higgs doublet model (2HDM) and parameter
constraints relevant for 
$\amu$. We employ a recent full two-loop computation of $\amu$ and 
take into account experimental constraints from Higgs and flavour
physics on the parameter space. Large $\amu$ is possible for light
pseudoscalar Higgs $\MA$ with large Yukawa couplings to leptons, and
it can be further increased by large $\MA$ coupling to top quarks. We
investigate in detail the maximum possible Yukawa couplings to leptons
and quarks of a light $\MA$, finding values of around $50\ldots100$
(leptons) and ${\cal O}(0.5)$ (quarks). As a result we find that an overall
maximum of $\amu$ in the 2HDM of more than $45\times10^{-10}$ is possible in a
very small parameter region around $\MMA=20$ GeV. The parameter
regions in which the currently observed deviation can be explained are
characterized.
\end{abstract}

\input{gm22HDM.sec1.tex}
\input{gm22HDM.sec2.tex}
\input{gm22HDM.sec3.tex}

\input{gm22HDM.sec4.tex}

\input{gm22HDM.sec5.tex}

\section*{Acknowledgments}
We gratefully acknowledge discussions with Jinsu Kim, Eung Jin Chun,
Wolfgang Mader, Mikolaj Misiak, and 
Rui Santos. 
The authors acknowledge financial support from DFG Grant
STO/876/6-1, and CAPES (Coordena\c{c}\~{a}o de Aperfei\c{c}oamento de
Pessoal de N\'{i}vel Superior), Brazil. The work has further been 
supported by the high-performance computing cluster Taurus at ZIH, TU
Dresden, and by the HARMONIA 
project under contract UMO-2015/18/M/ST2/00518 
(2016-2019).

\input{gm22HDM.app.tex}

\input{gm22HDM.biblio.tex}
\end{document}

%% file: gm22HDM.sec1.tex
\section{Introduction}

The two-Higgs doublet model (2HDM) is one of the most common
extensions of the Standard Model (SM). It is the simplest model with
non-minimal electroweak symmetry breaking, comprising two SU(2)
doublets and five physical Higgs bosons $\Mh$, $\MH$, $\MA$,
$\MHpm$, where $\Mh$ must be SM-like to agree with LHC-data. The
extra Higgs bosons are actively searched for at the 
LHC.

For more than a decade the measured value \cite{Bennett:2006fi} of the
anomalous magnetic 
moment of the muon $\amu=(g-2)_\mu/2$ has shown a persisting deviation
from the current SM prediction (for recent developments see
Refs.\ \cite{Kinoshita2012,Gnendiger:2013pva,Kataev:2012kn,SteinhauserQED}
(QED and electroweak corrections),
\cite{Davier,HMNT,Hagiwara:2017zod,KNTTalks,Jegerlehner:2017lbd,JegerlehnerSzafron,Benayoun:2012wc,Kurz:2014wya,Colangelo:2014qya,Colangelo:2014,Pauk:2014rfa,lattice,Ablikim:2015orh,Chakraborty:2015ugp}
(QCD corrections)). Using the evaluation of the indicated references,
the current deviation is
\begin{align}
a_\mu^{\text{Exp$-$SM}}&=
\begin{cases}
(26.8 \pm 7.6 ) \times 10^{-10} \mbox{\cite{Davier}}, \\
  (28.1 \pm 7.3) \times 10^{-10} \mbox{\cite{KNTTalks}},\\
  (31.3 \pm 7.7 ) \times 10^{-10} \mbox{\cite{Jegerlehner:2017lbd}}.
\end{cases}
\label{deviation}
\end{align}

 $\amu$ provides a tantalizing
hint for new physics. The hint might be strongly sharpened by a new
generation of $\amu$ measurements at Fermilab and J-PARC
\cite{Carey:2009zzb,Iinuma:2011zz}. Hence it is of high interest to
identify new physics models which are able to explain the current
deviation, or a future larger or smaller deviation.

Recently it has been repeatedly stressed that the 2HDM is such a model. This is a
non-trivial observation since the leading 2HDM contributions to $\amu$
arise only at the two-loop level and small Higgs masses are needed to
compensate the two-loop suppression. Specifically, Refs.\ \cite{ChunPassera,Wang:2014sda,Abe:2015oca,Crivellin:2015hha,ChunTakeuchi,ChunKim} have
studied the so-called type X (or lepton-specific) model,
Refs.\ \cite{Ilisie:2015tra,TaoHan,oldpaper} the more general
(flavour-)aligned model \cite{Pich:2009sp,pichtuzon}. In 
all these cases it was shown that a light pseudoscalar $\MA$ boson
with large couplings to leptons is viable and could explain
Eq.\ (\ref{deviation}) or at least most of it. Ref.\ \cite{TaoHan} has
also found an additional small parameter region with very light scalar
$\MH$; furthermore, Ref.\ \cite{Abe:2017jqo} has studied a $Z_4$-symmetric,
``muon-specific'' model which can explain Eq.\ (\ref{deviation}) for
$\tan\beta\sim1000$.

At the same time, the accuracy of the $\amu$ prediction in the 2HDM
has increased. Ref.\ \cite{oldpaper} has computed the 2HDM
contributions fully at the two-loop level, including all bosonic
contributions (from Feynman diagrams without closed fermion
loop). Prior to that, Ref.\ \cite{Ilisie:2015tra} had computed all
contributions of the Barr-Zee type \cite{BarrZee}. As a result of
these calculations, the 2HDM theory uncertainty is fully under control and
significantly below the theory uncertainty of the SM prediction and
the resolution of the future $\amu$ measurements.

Here we employ the full two-loop prediction to carry out a detailed
phenomenological study of $\amu$ in the general flavour-aligned 2HDM
and of the parameters 
relevant for $\amu$. In detail, the questions we consider are
\begin{itemize}
  \item What are the constraints on the 2HDM parameters most relevant
    for $\amu$ (the
    mass of the $\MA$ boson and its Yukawa couplings to leptons and
    quarks, and further 2HDM masses and Higgs potential parameters)?
    \item In which parameter region can the 2HDM accommodate the
      current deviation in $\amu$ (or a future, possibly larger or
      smaller deviation)?
      \item What is the overall maximum possible value of $\amu$ that
        can be obtained in the 2HDM (for various choices of restrictions on the
        Yukawa couplings)?
\end{itemize}
We will generally focus on the promising scenario with $\MMA<\MMh$ and
allow for general flavour-aligned Yukawa couplings but will comment
also on the more restrictive case of the lepton-specific type X
model. We will take into account constraints from theoretical
considerations such as tree-level unitarity and perturbativity,
experimental constraints from collider data from LHC and LEP, and
constraints from B- and $\tau$-physics.

The outline of the paper is as follows. In section 2, we describe our
setup and give details on the definition of the 2HDM. Section 3 then
discusses the detailed constraints on the parameters most relevant for
$\amu$ in the 2HDM: on the Higgs masses, on the Yukawa
couplings, and on Higgs potential parameters and Higgs self
couplings. Section 4 gives an updated discussion of the full bosonic 
two-loop contributions, taking into account detailed constraints on
the parameters. Section 5 finally gives the results on $\amu$ in the
2HDM. The results are presented both as contour plots in parameter
planes, and as plots showing the maximum possible values of $\amu$ in
the 2HDM.

%% file: gm22HDM.sec2.tex
\section{Setup}

In this section we provide the basic relations for the two-Higgs doublet model (\thdm) and describe our technical setup.

\subsection{Definition of the \thdm}

We use the 2HDM with general Higgs potential in the notation of
Ref.~\cite{Gunion:2002zf,Branco:2011iw}, 
\begin{align}\label{potential}
V\left(\phi_{1},\phi_{2}\right) &= m^{2}_{11}\phi^{\dagger}_{1}\phi_{1}+m^{2}_{22}\phi^{\dagger}_{2}\phi_{2}
- \{m^{2}_{12}\phi^{\dagger}_{1}\phi_{2} + {\text{H.c.}}\}\nonumber\\
&+\frac{\lambda_{1}}{2}(\phi^{\dagger}_{1}\phi_{1})^{2}
+\frac{\lambda_{2}}{2}(\phi^{\dagger}_{2}\phi_{2})^{2}
+\lambda_{3}(\phi^{\dagger}_{1}\phi_{1})(\phi^{\dagger}_{2}\phi_{2})\nonumber\\
&+\lambda_{4}(\phi^{\dagger}_{1}\phi_{2})(\phi^{\dagger}_{2}\phi_{1})
+\frac{1}{2}\left\{\lambda_{5}(\phi^{\dagger}_{1}\phi_{2})^{2}+{\text{H.c.}}\right\}\nonumber\\
&+\left\{[\lambda_{6} (\phi^{\dagger}_{1}\phi_{1}) + \lambda _{7} (\phi^{\dagger}_{2}\phi_{2})]\phi^{\dagger}_{1}\phi_{2} + {\text{H.c.}}\right\}.
\end{align}

In the usual type I, II, X, Y models a $Z_2$ symmetry is assumed which enforces the two parameters $\lambda_6$ and $\lambda_7$ to vanish. In the
following we will investigate both the case with
$\lambda_6=\lambda_7=0$ and the case with non-vanishing $\lambda_6, \lambda_7$. Since
we focus on the muon magnetic moment, which is not enhanced by CP
violation, we assume all parameters to be real\footnote{See footnote 5 in Section~\ref{sec:conclusions}.}. In the minimum of the
potential the two Higgs  
doublets acquire
the vacuum expectation values (VEVs) $v_{1,2}$ with the ratio
$\tan\beta=v_2/v_1$. It is
then instructive to rotate the doublets by the angle $\beta$ to the
so-called  
Higgs basis~\cite{Gunion:2002zf}, in which one doublet has the full
SM-like VEV $v=\sqrt{v_1^2+v_2^2}$ and the other doublet has zero
VEV. The second doublet then contains the physical CP-odd Higgs $\MA$
and the charged  
Higgs $\MHpm$, and 
the physical CP-even Higgs fields $h,H$ correspond to mixtures between the two doublets in the Higgs basis with mixing angle $(\alpha-\beta)$.
In practice we will choose the following set of independent input parameters:
\begin{align}
M_{h,H,A,H^\pm},\tan\beta,\CBA,\lambda_1,\lambda_6,\lambda_7,
\label{parameterlist}
\end{align}
where $\CBA\equiv\cos(\beta-\alpha)$ and similar for $\SBA$.
We will further choose $\Mh$ to be the approximately SM-like Higgs
state, which means that  the mass $\MMh$ is fixed to the observed
value of $125$ GeV and that the mixing angle $\CBA$ is 
small. 
It should be noted that all  parameters in this list enter the
prediction of the muon $g-2$ only at the two-loop level and hence do
not have to be renormalized.  

For the Yukawa couplings we choose the setup of the (flavour-)aligned 2HDM of Ref.\ \cite{Pich:2009sp}. In this setup one assumes the following structure of the
Yukawa couplings in the Higgs basis: the SM-like doublet has SM-like Yukawa couplings by construction; the other doublet has couplings proportional to the SM-like ones,
with proportionality factors $\zeta_l$ (for charged leptons), $\zeta_{u,d}$ (for up- and down-type quarks). For the mass-eigenstate Higgs bosons this
implies the following Yukawa Lagrangian:
\begin{align}\label{yukawaalign}
{\cal{L}}_Y =& -\frac{\sqrt{2}}{v} H ^+ \Big({\bar{u}}[\zeta_d V_{\text{CKM}} M_d P _{\text{R}} - \zeta_u M_u V_{\text{CKM}} P _{\text{L}}] d + \zeta_l {\bar{\nu}} M_l P _{\text{R}} l \Big)\nonumber\\
  &- \sum _{{{\cal S}=\Mh,\MH,\MA}}\sum_{{f=u,d,l}}\,{\cal S}{\bar{f}} y _f ^{\cal S} P _{\text{R}} f + {\text{H.c.}},
\end{align}
where $P_{{\text{R}},{\text{L}}} = \frac{1}{2}(1 \pm \gamma _5)$, and $V_{\text{CKM}}$ is the Cabibbo-Kobayashi-Maskawa matrix. $M_f$ denotes the diagonal $3\times3$ mass matrices. The Yukawa coupling matrices are defined as 
\begin{align}\label{yukawas}
y_{f}^{\cal S}&=\frac{Y_{f}^{\cal S}}{v} M_f, 
\end{align}
where
\begin{align}\label{yukawacs}
Y^{\Mh}_{f} &= \SBA+\CBA\zeta_{f}, \nonumber\\
Y^{\MH}_{f} &= \CBA-\SBA\zeta_{f}, \nonumber\\
Y^{\MA}_{d,l} &= i\zeta_{d,l}, \nonumber\\
Y^{\MA}_{u} &= -i \zeta_{u}. 
\end{align}
    
The flavour-aligned 2HDM contains the usual type I, II, X, Y models as special cases, see table \ref{table:yukXYZ}. Most notably, in type II, the product 
$|\zeta_u \zeta_d|=\cot\beta\; \tan\beta=1$ is never small, implying very strong constraints from $b\to s\gamma$ for all values of 
$\tan\beta$ \cite{MisiakSteinhauser}. And in type X, $\zeta_l=-\tan\beta$ and $\zeta_u=\zeta_d=\cot\beta$ cannot be simultaneously large.

\begin{table}[t]
\begin{center}
\begin{tabular}{lrrrr}
\hline\hline
           &\qquad Type I      &\qquad Type II      &\qquad Type X       &\qquad Type Y        \\ \hline        
$\zeta_{u}$ &\qquad $\cot\beta$ &\qquad $\cot\beta$  &\qquad $\cot\beta$  &\qquad $\cot\beta$  \\ \hline
$\zeta_{d}$ &\qquad $\cot\beta$ &\qquad $-\tan\beta$ &\qquad $\cot\beta$  &\qquad $-\tan\beta$ \\ \hline
$\zeta_{l}$ &\qquad $\cot\beta$ &\qquad  $-\tan\beta$ &\qquad $-\tan\beta$ &\qquad $\cot\beta$  \\ \hline\hline
\end{tabular}
\end{center}
\caption{Relation between the Yukawa parameters $\zeta_{f}$ in the
  general, aligned 2HDM and the usual type I, II, X, and Y models.}
\label{table:yukXYZ}
\end{table}

As shown in Refs.\ \cite{pichtuzon,Gori:2017qwg} the flavour-aligned
scenario is minimal flavour violating and even though the alignment is
not strictly protected by a symmetry, it is  
numerically rather stable under renormalization-group running. Hence
we regard it as a theoretically and phenomenologically well motivated
and very general scenario. 

\subsection{Technical  remarks}
\label{sec:technicalremarks}

In order to check the viability of parameter points against
experimental and theoretical constraints, we have adopted the routines
implemented in the 2HDMC code\,\cite{2HDMC}, which allows checks
regarding theoretical constraints such as stability, unitarity, and
perturbativity of the quartic couplings; the S, T, U precision
electroweak parameters; and data from colliders implemented in the
HiggsBounds and HiggsSignals packages\,\cite{HB,HS}. 

For our later scans of parameter space we started with a wide range of
all Higgs potential parameters in Eq.\ (\ref{parameterlist}) and the
Yukawa parameters $\zeta_{l,u,d}$. This range was narrowed down to
\begin{align}
  \zd&=-0.7\ldots1.1, & \lambda_{1}&=0\ldots2\pi,
  \nonumber\\
  \lambda_{6}&=-2\ldots2,&
  \lambda_{7}&=-3\ldots3,
\label{parameterranges}
  \\
  \tan\beta&=0.3\ldots2,&|\CBA|&<1/|\zl|,\nonumber
\end{align}
after checking that this covers the parameter space with the largest
possible contributions to all quantities of interest. Unless specified
differently, these are the 
parameter ranges used in our scatter plots. In the plots evaluating
$\amu$, in addition we set $\zd=0$ to be specific, because this
parameter has a very small influence on $\amu$.

Regarding statistics, we have adopted the following procedure: first
we constructed a $\chi^2$ distribution for the physical process under
consideration, and then computed its respective p-value distribution,
assuming that the errors are gaussian and robust as usual
\cite{PDG}. Finally, we required that the p-value for the considered
observable (or set of observables) is greater than 0.05 (corresponding
to a 95\% CL region). For
the constraints to be discussed in Sec.\ref{sec:zlconstraints}, this approach is
slightly different from the one implemented in Ref.\ \cite{ChunKim},
but we checked that the resulting exclusion contours are very similar.

%% file: gm22HDM.sec3.tex
\section{Constraints}
\label{sec:constraints}

In this section we provide a detailed investigation of experimental
constraints on the 2HDM parameter space with general flavour-aligned
Yukawa couplings. Earlier studies  
 \cite{ChunPassera,Wang:2014sda,Ilisie:2015tra,Abe:2015oca,Crivellin:2015hha,ChunTakeuchi,TaoHan,ChunKim,oldpaper}
 and our later considerations show that $\amu$ can be promisingly
 large for small $\MMA$ and large $\zeta_l$ and $\zeta_u$, so we
 focus on this scenario.  
Our study can be regarded as a 
generalization of Refs.\ \cite{ChunPassera,ChunKim,Abe:2015oca}, which focused on the lepton-specific (type X) case, where $\zeta_u=-1/\zeta_l=1/\tan\beta$, and as 
complementary to Ref.\ \cite{TaoHan}, which focused 
on correlations in scans of parameter space. Our questions are: what
are the maximum values of $\zeta_l$ and $\zeta_u$ and other relevant
parameters, and how do these maximum values depend on the 
value of the small Higgs mass $\MMA$ or the heavy Higgs masses?

We will begin with the most direct and basic constraints on the
scenario with small $\MMA$ from collider physics, then focus on
maximum possible values of $\zeta_l$ and $\zeta_u$ and correlated
parameters. 
\input{gm22HDM.sec31.tex}
\input{gm22HDM.sec32.tex}
\input{gm22HDM.sec33.tex}

%% file: gm22HDM.sec31.tex
\subsection{Basic collider constraints on small $\MMA$ and on mixing angle $\cos(\beta-\alpha)$}
\label{sec:basicconstraints}
The scenario with light CP-odd Higgs boson $\MA$ is obviously strongly constrained by collider physics. The most immediate constraints arise from 
negative results of direct $\MA$ searches. On the one hand these results
imply upper limits of the couplings between the $\MA$ and $W$ and $Z$
bosons and thus on the mixing angle 
$\CBA$. However, below we will find much more severe limits on $\CBA$, which are specific to our scenario with large $\zeta_l$, so we will
discuss only those in detail. On the other hand the negative searches for $\MA$ imply upper limits on $\zeta_u$ in a restricted range of $\MA$ masses; we will discuss 
these in subsection
\ref{sec:zetauconstraints}.

In the remainder of this subsection we will discuss more interesting
collider constraints on our scenario, 
which arise from measurements of the decays of the observed SM-like
Higgs boson at the LHC. First, the LHC measurements of/searches for
SM-like Higgs decays into $\tau$ pairs or muon pairs imply
limits on the coupling of the SM-like Higgs boson to $\tau$-leptons and
muons. Expressed in terms of signal strengths, the recent
Refs.\ \cite{Sirunyan:2017khh,Aaboud:2017ojs} obtain
\begin{align}
\mu_{\tau} &= 1.09 ^{+0.27}_{- 0.26}, \\
\mu_{\mu} &= -0.1 \pm 1.4,
\end{align}
implying that the effective coupling of the SM-like Higgs to leptons
$Y^{\Mh}_{l}$ in Eq.\ (\ref{yukawacs}) cannot deviate strongly from
unity and thus,\footnote{In case of the wrong-sign Yukawa limit, see
  below, the l.h.s.\ is exactly 2. Still, the approximate form of
  Eq.\ (\ref{yhleptonconstraint}) holds in this case.
}
\begin{align}
|\CBA\zeta_l|<{\cal O}(1),
  \label{yhleptonconstraint}
\end{align}
The approximate form of this relation is sufficient for our
purposes. The important points are that ($i$)
for a given, large $\zeta_l$, the mixing angle $\CBA$ is strongly constrained particularly by the $\tau$-coupling to be at most of the order of a 
percent, and ($ii$) the product $\CBA\zeta_l$ cannot
constitute an enhancement factor.

A second important implication of the SM-like Higgs decay measurements comes from the decay mode $h\to \MA\MA$, which is possible if $\MMA<\MMh/2$. A 
significant  branching fraction
for this decay is excluded by the agreement of the observed Higgs decays with the SM predictions. This implies strong constraints on the corresponding triple 
Higgs coupling $C_{hAA}$, given explicitly in Eq.\ (\ref{eq:hAA}) in
the Appendix. 
It is therefore illuminating to analyze analytically the conditions for vanishing coupling $C_{hAA}$. We have to distinguish two cases:
\begin{itemize}
\item $\MMA<\MMh/2$ and large $\tan\beta$: In this limit, the
  requirement $C_{hAA}=0$ reduces to
\begin{align}
\CBA=2/\tan\beta+{\cal O}(1/\tan^2\beta)
\label{etasolution}
\end{align}
For the type X model, where $\tan\beta=-\zeta_l$, this and
Eq.\  (\ref{yukawas}) implies
$Y^{\Mh}_l\approx-1$, the so-called wrong-sign muon Yukawa coupling,
discussed recently in Ref.\ \cite{ChunTakeuchi}. In 
the general case, this relation, together with the limit on $\CBA$
from Eq.\ (\ref{yhleptonconstraint}), implies a lower limit on
$\tan\beta$, which is  
of the form $\tan\beta\gg|\zeta_l|$. This parameter region does not
lead to distinctive phenomenology; we will not discuss it further.
\item $\MMA<\MMh/2$ and small $\tan\beta$: In this case, one can solve
  the requirement $C_{hAA}=0$ for $\lambda_1$. The exact solution can
  be read off from Eq.\ (\ref{eq:hAA}). We provide the solution here
  for $\CBA = 0$,
\begin{align}
\lambda_1 &=\frac{\MMh ^2}{v ^2}\left(1 - \frac{\tb ^2}{2}\right) + \left(\frac{\MMH ^2-\MMA ^2}{v ^2}\right)\tb ^2 - \frac{3}{2}\lambda _6 \tb + \frac{1}{2}\lambda _7 \tb ^3. 
\label{lambda1solution}
\end{align}
\end{itemize}
We checked that even if we allow $C_{hAA}\ne0$, no significant
deviations from relations (\ref{etasolution}) or
(\ref{lambda1solution}) are experimentally allowed if $\MMA<\MMh/2$. 
Hence we will always impose these relations exactly and fix either
$\CBA$ or $\lambda_1$ in terms of these relations if
$\MMA<\MMh/2$. 

%% file: gm22HDM.sec32.tex
\subsection{Constraints on the lepton Yukawa coupling $\zl$}
\label{sec:zlconstraints}

Next we present the upper limits on $|\zl|$, the lepton Yukawa
coupling parameter in the flavour-aligned 2HDM. This
parameter governs in particular the couplings $Y^{\MA}_{l}$ of
$\MA$ to $\tau$-leptons or muons. After earlier similar studies in
Ref.~\cite{Abe:2015oca}, precise limits on $\zl$ have been obtained in
Ref.~\cite{ChunKim} for the case of the type X model, where
$\zl=-\tan\beta$. We have repeated 
the analysis for the case of the flavour-aligned model, finding
essentially the same upper limit on $|\zl|$ as Ref.~\cite{ChunKim}
finds on $\tan\beta$ (except at small $\MMA$ due to additional
collider constraints, see below).\footnote{%
Small differences also arise due to our slightly different treatment of
  the statistical significances.}

\begin{figure}
\centering{    \includegraphics[scale=.9]{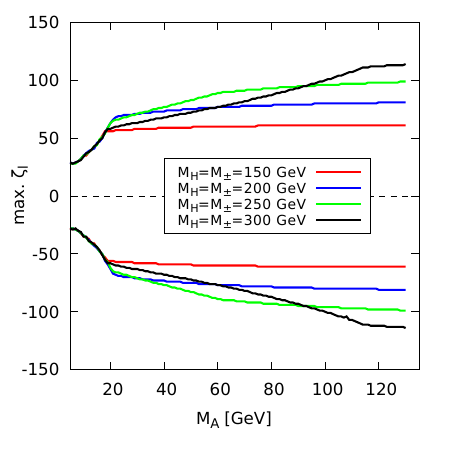}}
\caption{Maximum possible values of the lepton Yukawa parameter $\zl$,
given constraints from $\tau$- and $Z$-decays and collider data, as a
function of $\MMA$ for several values of $\MMH=\MMHpm$ as indicated. }
\label{fig:maximumzl}
\end{figure}

The upper limits on $|\zl|$ arise on the one hand from experimental
constraints on 
the $\tau$-decay mode $\tau\to\mu\nu_\tau
\bar\nu_\mu$ versus other decay modes and on leptonic $Z$-boson
decays.  2HDM diagrams contributing 
to these decays involve tree-level or loop exchange of $\MA$ or
$H^\pm$. They are enhanced by $\zl$ and lead to disagreement with
observations if $|\zl|$ is too large. We computed the $\tau$- and
$Z$-boson decays and the $\Delta\chi^2$ corresponding to the deviation
from experiment as described in Ref.~\cite{ChunKim} and
sec.\ \ref{sec:technicalremarks}.

On the other hand, further constraints on $\zl$ arise from collider
data.
In particular,  for small $\MMA$ ($5<\MMA<20$
GeV) the upper bound of $|\zl|$ is dominated by the LEP process 
$ee\to \tau\tau(\MA)\to \tau\tau(\tau\tau)$ which was probed by the
DELPHI collaboration \cite{Abdallah:2004wy}. In this decay, the
electron positron pair annihilates into a $Z$-boson which further
generates a pair of $\tau$-leptons. From one of those, a short-lived $\MA$ boson
is created in resonance, producing finally two more taus.

Our resulting upper limits on $|\zl|$ are shown in
Fig.~\ref{fig:maximumzl} as functions of $\MMA$ for various choices of
$\MMHpm$. The limits are generally between $|\zl|<40$ and
$|\zl|<100$. In most of the parameter space the limits are
dominated by the $\tau$-decay constraints, which become weaker for
larger $\MMA$ and 
larger $\MMH,\MMHpm$. The constraints from $Z$-boson decays become dominant
for heavy Higgs masses above around 250 GeV. For even higher Higgs
masses, these limits reduce the maximum $|\zl|$ (see the black lines in
Fig.\ \ref{fig:maximumzl}). Aiming for largest
possible Yukawa 
couplings, the $Z$-boson decay constraints imply that even larger
heavy Higgs masses will not help.
The constraints from LEP data are dominant for small $\MMA<20$ GeV and
significantly reduce the maximum $|\zl|$ in this parameter region.

%% file: gm22HDM.sec33.tex
\subsection{Constraints on the up-type Yukawa coupling $\zeta_u$}
\label{sec:zetauconstraints}

In this subsection we present the upper limits on $\zeta_u$, the
parameter for up-type quark Yukawa couplings. This is a central part
of our analysis, showing characteristic differences between the case
of the type X model and the general flavour-aligned model.
In what follows, we will focus on  negative $\zl$
(like in the type X model where $\zl=-\tan\beta$) and positive $\zu$,
which leads to larger contributions to $\amu$.

\begin{figure}
  \begin{center}
    \null\quad
    \begin{subfigure}[c]{.45\textwidth}
      \label{fig:bmumu}
      \begin{center}
        \includegraphics[scale=.55]{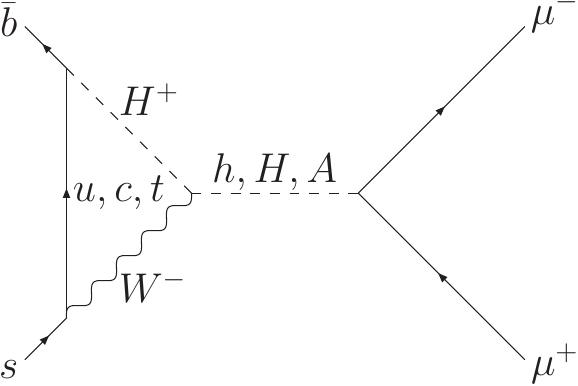}
      \end{center}
      \subcaption{}
    \end{subfigure}
    \quad
    \begin{subfigure}[c]{.45\textwidth}
      \label{fig:bsg}
      \begin{center}
        \includegraphics[scale=.55]{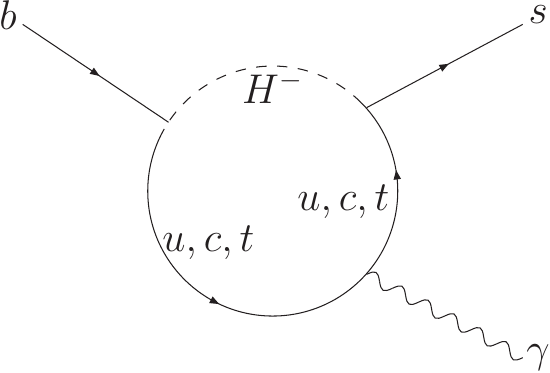}
      \end{center}
      \subcaption{}
      \end{subfigure}
  \end{center}
  \caption{\label{fig:bdiagram} Sample Feynman diagrams for the
    processes  $B_s\to\mu^+\mu^-$ and $b\to s\gamma$, which depend on the
    Yukawa couplings of up- and down-type quarks and leptons.}
\end{figure}

In type II or type X models $\zeta_u$ is always small for large
lepton Yukawa coupling, because
$\zeta_u=-1/\zeta_l=1/\tan\beta$. However, if general Yukawa
couplings are allowed, $\zeta_u$ can be larger. The maximum possible
value is interesting not only for $g-2$ but also in view of future LHC
searches for a low-mass $\MA$.

We find that $\zeta_u$, in the scenario of $\MMA<\MMh$ and large
$\zeta_l$, is constrained in a complementary way by B-physics on the
one hand, and by LHC-data on the other hand.

Beginning with B-physics, the most constraining observables for this
scenario are $b\to s\gamma$ and $B_s\to\mu^+\mu^-$. The sample diagrams
shown in Fig.~\ref{fig:bdiagram} illustrate that the $\thdm$ predictions
depend on combinations of all Yukawa parameters $\zeta_l$, $\zeta_u$,
$\zeta_d$ and on the Higgs masses $\MMA$ and $\MMHpm$. We have
implemented the analytical results for the predictions presented in
Refs.~\cite{pichetal,watanabe} (Ref.~\cite{watanabe} has also
considered further observables, which however do not constrain the
parameter space further; see also Ref.~\cite{Arnan:2017lxi} for improvements on the precision of B-physics observables).

\begin{figure}
  \begin{center}
    \begin{subfigure}[l]{.48\textwidth}
      \label{fig:plotzUzD40}
      \includegraphics[scale=.9]{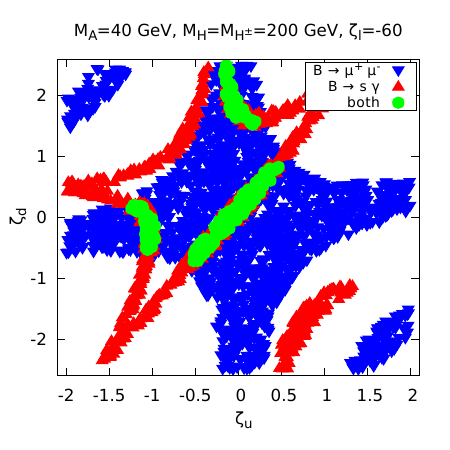}
      \subcaption{}
    \end{subfigure}
    \begin{subfigure}[r]{.48\textwidth}
      \label{fig:plotzUzD50}
      \includegraphics[scale=.9]{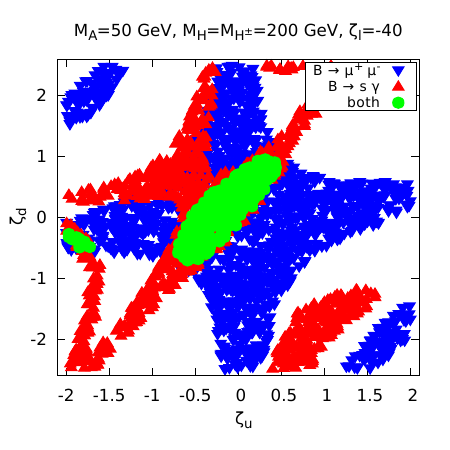}
      \subcaption{}
      \end{subfigure}
  \end{center}
  \caption{Allowed parameter regions in the $\zu$--$\zd$-plane given
    constraints from $b\to s \gamma$ or $B_s\to\mu^+\mu^-$ or the
    combination. The parameters are chosen as indicated. }
  \label{fig:plotzUzD}
\end{figure}

To illustrate the interplay between the observables we show first
Fig.~\ref{fig:plotzUzD}.
It shows the $2\sigma$ regions in the
$\zeta_u$--$\zeta_d$-plane allowed by either
$b\to s \gamma$ or $B_s\to\mu^+\mu^-$ alone or by the combination.
In the figure, the representative values 
$\MMHpm=200\text{ GeV, and }(\MMA,\zeta_l)=(40\text{ GeV, }-60)\text{ or }(50\text{ GeV, }-40)$ are fixed, as indicated. 

Both observables on their own would allow values of $\zeta_u\gg1$, by
fine-tuning $\zeta_d$ and $\zeta_u$. However, the combination of both
observables implies an upper limit on $\zeta_u$, which in this case is
$\zeta_u < 0.5$.
\footnote{%
  For some values of $\MMA$, $\MMHpm$, separate ``islands'' in the
  $\zeta_u$--$\zeta_d$-plane at higher $\zeta_u$ can be allowed. They
  can be excluded by the universal bound $|\zeta_u|<1.2$ derived from $R_b$ in
  Ref.~\cite{pichtuzon}, and by the similar bound derived from $\Delta
  M_s$ in Ref.\ \cite{watanabe}.}

\begin{figure}
  \begin{subfigure}[]{.5\textwidth}
    \includegraphics[scale=.9]{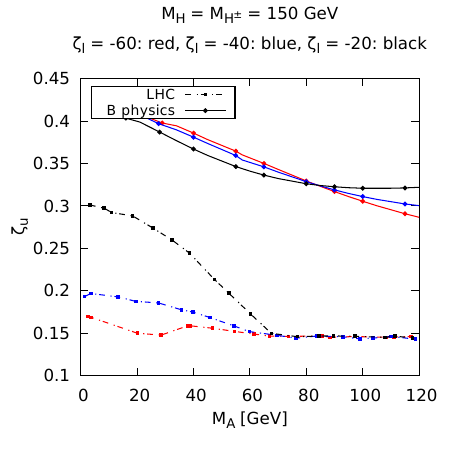}
    \subcaption{\label{fig:LHCB150}}
  \end{subfigure}
   \begin{subfigure}[]{.5\textwidth}
     \includegraphics[scale=.9]{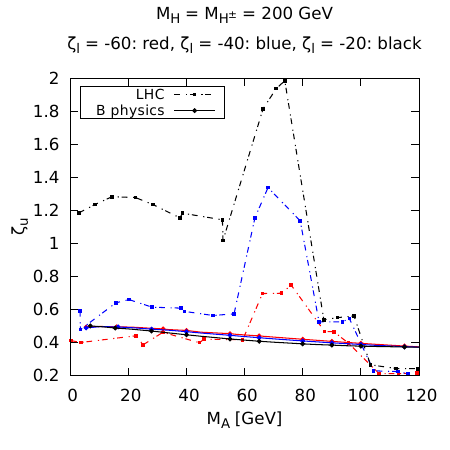}
     \subcaption{\label{fig:LHCB200}}
   \end{subfigure}\vspace{1cm}\\
  \begin{subfigure}[]{.5\textwidth}
    \includegraphics[scale=.9]{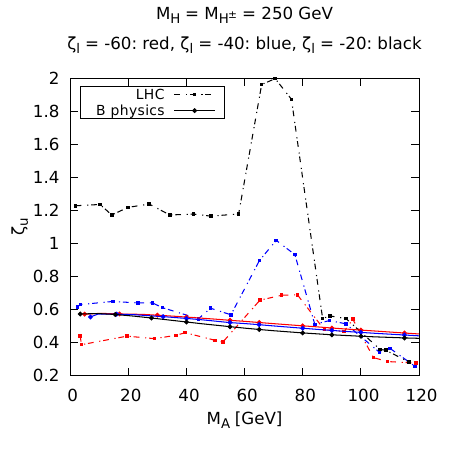}
    \subcaption{\label{fig:LHCB250}}
  \end{subfigure}
  \begin{subfigure}[]{.5\textwidth}
    \includegraphics[scale=.9]{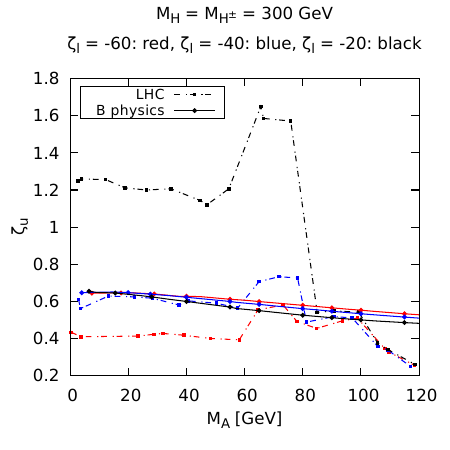}
    \subcaption{\label{fig:LHCB300}}
  \end{subfigure}
\caption{\label{fig:LHCBmaximumplots} The maximum allowed values of
  $\zeta_u$ as function of $\MMA$, for different values of $\MMH,
  \MMHpm$ and $\zeta_l$ as indicated. The continuous lines correspond
  to the upper limit derived from B-physics alone, the dashed lines to
  the upper limit derived from LHC-Higgs physics alone. }
\end{figure}

By performing a similar analysis repeatedly, we obtain maximum values
of $\zeta_u$ as function of $\MMA$, $\MMHpm$ and $\zeta_l$. The result
will be shown below in the plots of Fig.~\ref{fig:LHCBmaximumplots} as continuous lines.
Each solid line corresponds to the maximum
allowed value (by B-physics) of $\zeta_u$, as a function of $\MMA$ and
for fixed values of $\MMHpm$ and $\zeta_l$. The dependence on $\MMA$,
$\MMHpm$ and $\zeta_l$ is mild. Generally, the upper limit on
$\zeta_u$ is between $0.3$ and $0.6$.

\begin{figure}[ht]
  \begin{subfigure}[]{.5\textwidth}
    \includegraphics[scale=.9]{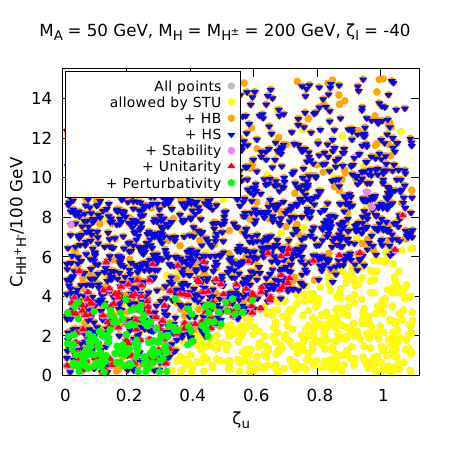}
  \subcaption{}
  \label{fig:zuscattera}
\end{subfigure}
  \begin{subfigure}[]{.5\textwidth}
\includegraphics[scale=.9]{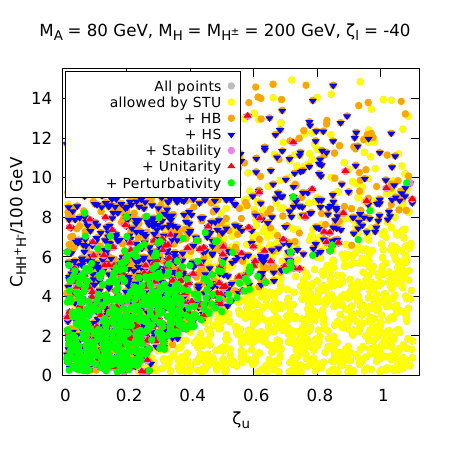} 
  \subcaption{}
  \label{fig:zuscatterb}
\end{subfigure}
\caption{ Allowed ranges of $\zu$ and the triple Higgs coupling
  $C_{HH^+H^-}$, given certain constraints, see legend and text. The
  constraints are applied successively.  The scanned
  parameter space is defined by Eq.\ (\ref{parameterranges}), with
  Eq.\ (\ref{lambda1solution}) in case $\MMA<\MMh/2$. }
\label{fig:zuscatter}
\end{figure}

Turning to LHC-Higgs physics, the dashed lines in the plots of
Fig.~\ref{fig:LHCBmaximumplots} show the maximum $\zeta_u$ allowed by
LHC collider constraints. These constraints on $\zeta_u$ arise from
several processes and measurements:
\begin{itemize}
\item
  $pp\to \MA \to \tau\tau$ for $\MMA>\MZ$
  \cite{CMS:2015mca}. In our scenario $\MA$ decays essentially to $100\%$ into
  $\tau\tau$. Hence the measurement constrains the production rate of
  $\MA$, which proceeds via top-quark loop and gluon fusion and is thus
  governed by $\zeta_u$. Hence this measurement provides an
  essentially universal upper
  limit of approximately $\zeta_u<0.2$ which becomes valid above $\MMA>100$~GeV.
\item
  $pp\to H\to\tau\tau$ \cite{CMS:2015mca} if $H\to \MA\MA$ is kinematically
  forbidden. Similar to the previous case, $\MH$ is produced in gluon
  fusion via a top-loop, so its production rate is governed by
  $\zeta_u$; it decays essentially always into a $\tau$-pair. Hence,
  again, this measurement places an essentially universal upper limit
  on $\zeta_u$, valid if $\MMA>\MMH/2$. In the plots, this limit can
  be seen for $\MMH=150$~GeV and $\MMA>75$~GeV.
\item
  $pp\to H\to\tau\tau$ \cite{CMS:2015mca} if $H\to \MA\MA$ is kinematically
  allowed. This case is relevant in the largest region of parameter
  space, including the regions with the peak structures in which the collider limits become
  rather weak and $\zl$-dependent. The scalar Higgs $\MH$ is
  produced in gluon fusion via a top-loop, so its production rate is
  governed by $\zeta_u$; its two most important decay modes are $\MH\to\MA\MA$
  and $\MH\to \tau\tau$. Hence, the signal
  strength for the full process depends not only on $\zeta_u$ but also on the
  triple Higgs coupling $C_{HAA}$, which is strongly correlated with
  $C_{HH^+H^-}$ given in Eq.~(\ref{eq:CHHpHm}). The signal strength can be
  suppressed by  small $\zeta_u$ (which suppresses the production) or
  by large $C_{HH^+H^-}$ (which suppresses the decay to
  $\tau\tau$).

  Hence we show the allowed ranges of $\zeta_u$ and the triple Higgs
  coupling $C_{HH^+H^-}$ in Fig.~\ref{fig:zuscatter}, for the
  representative values $\MMA=50/80$~GeV, $\MMH=\MMHpm=200$~GeV,
  $\zeta_l=-40$. The
  colours indicate the successive application of constraints from
  the electroweak S, T, U parameters, HiggsBounds, HiggsSignals,
  and tree-level stability, unitarity and perturbativity (as
  implemented in 2HDMC \cite{2HDMC}). The border of the yellow region
  shows clearly the correlation between the two couplings mentioned
  above, needed to evade  the constraints from   $pp\to H\to\tau\tau$
  searches. The larger the triple Higgs coupling, the larger $\zeta_u$
  can be. However, perturbativity restricts the triple Higgs coupling,
  and this restriction depends on whether   $\MMA<\MMh/2$ holds or
  not. If   $\MMA<\MMh/2$, the relation
  (\ref{lambda1solution}) following from setting to zero
  Eq.\ (\ref{eq:hAA}) has to be used, and the maximum triple Higgs
  coupling and thus the maximum $\zu$ is smaller.

  As a result of this combination of constraints, the LHC-Higgs limits
  on $\zeta_u$ are rather loose for $\MMA$ between $\MMh/2$ and around
  $M_Z$ (explaining the peaks in Fig. \ref{fig:LHCBmaximumplots}), and
  stronger for lower $\MMA$. The precise value of the 
  limits depends on $\zeta_l$, which also influences the branching
  ratio $\MH\to\tau\tau$. 
\item
  We also mention the analysis of Ref.~\cite{ChunTakeuchi}, where
  LHC-constraints on the type X model have 
  been studied; since $\zu$ is negligible in the type X model, those
  constraints are weaker than the ones we consider here, and they do
  not limit $\zu$. Still, that analysis shows that data from
  multi-Higgs production followed by 
  decays into multi-$\tau$ final states leads to interesting (mild)
  constraints on heavy $\MMH$, $\MMHpm$.
\end{itemize}

%% file: gm22HDM.sec4.tex
\section{Bosonic contributions to $\amu$ and relevant parameter constraints}
\label{sec:amub}

As discussed in the previous section, the 2HDM parameter region of interest
for $\amu$ is characterized by large Yukawa coupling parameter $\zl$
and small pseudoscalar mass $\MMA$. The bosonic two-loop contributions
$\amub$ computed in Ref.\ \cite{oldpaper} depend on a large number of
additional parameters: the 
physical Higgs masses $\MMH$, $\MMHpm$, the mixing angle
$\CBA$, $\tan\beta$, and the Higgs potential parameters
$\lambda_1$ and $\lambda_{6,7}$. In the present section we provide an
overview of the influence of these parameters, constraints on their
values, and update the analysis of Ref.\ \cite{oldpaper} given those
constraints. As a result we derive the maximum possible values of the
bosonic two-loop contributions to $\amu$.

The bosonic two-loop contributions can be split into three parts
\cite{oldpaper},
\begin{align}
\amub &= \amu ^{\text{EW add.}} + \az + \amu ^{\text{Yuk}},
\end{align}
where $\amu^{\text{EW add.}}$ denotes the difference between the
contribution of the SM-like Higgs in the 2HDM and its SM counterpart;
$\az$  and $\amu ^{\text{Yuk}}$ denote remaining bosonic contributions
without/with Yukawa couplings.

We begin with a discussion of $\amu ^{\text{EW add.}}$, which is
approximately given by $\amu ^{\text{EW add.}} = 2.3 \times 10^{-11}
\,\CBA\,\zl$.  As discussed in section
\ref{sec:basicconstraints}, the product $\CBA\,\zl$ is
restricted by Higgs signal strength measurements to be smaller than
unity. Hence this product can never be an enhancement
factor. Specifically, as a result we obtain the conservative limit
\begin{align}
|\amu ^{\text{EW add.}}| &< 0.2\times10^{-10}\,,
\label{amuewadd}
\end{align}
such that these contributions are negligible.

Next we consider $\az$, the contribution from diagrams in which the
extra 2HDM Higgs bosons couple only to SM gauge bosons and not to
fermions. Similar to the quantity $\Delta\rho$, this contribution is
enhanced by large mass splittings $|\MMH-\MMHpm|$ between the heavy
Higgs bosons. Conversely, constraints on $\Delta\rho$ restrict this
mass splitting \cite{ChunPassera,Hessenberger:2016atw} and thus
$\az$. We find that $\az$ is similarly negligible as
Eq.\ (\ref{amuewadd}).

Finally we turn to $\amu ^{\text{Yuk}}$, the potentially largest
bosonic two-loop contribution. Ref.\ \cite{oldpaper} has decomposed
this contribution into several further subcontributions depending on
the appearance of triple Higgs couplings, the mixing angle
$\CBA$ and the Yukawa parameter $\zl$. Among these
parameters, the product $\CBA\,\zl$ is restricted as
discussed above; furthermore, the triple Higgs couplings are
constrained by perturbativity. Inspection of the results of
Figs.\ 5 and 6 of Ref.\ \cite{oldpaper} then shows that all
subcontributions to  $\amu ^{\text{Yuk}}$ are at most of the order
$10^{-11}$, with the exception of the ones enhanced by the
triple Higgs coupling $C_{HH^+H^-}$.

\begin{figure}
  \begin{subfigure}[]{.5\textwidth}
    \includegraphics[scale=.9]{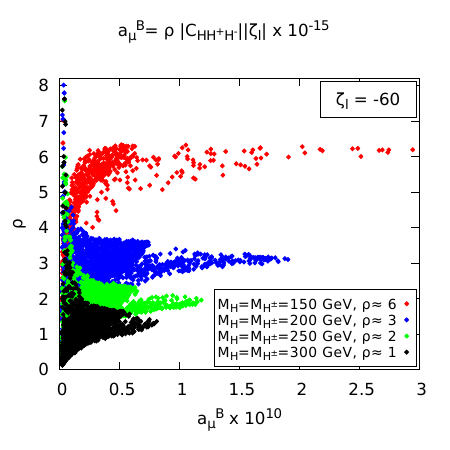}
    \subcaption{\label{fig:rho}}
  \end{subfigure}\quad
  \begin{subfigure}[]{.5\textwidth}
    \includegraphics[scale=.9]{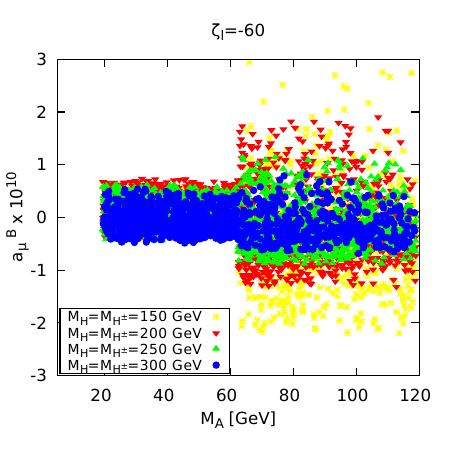}
    \subcaption{\label{fig:multiMamuB}}
  \end{subfigure}\quad
  \caption{The bosonic contributions $\amub$. (a)
  The proportionality factor $\rho$ defined in Eq.\ (\ref{rho}) for a
  scan of parameter space with different values of the heavy Higgs
  masses. (b) The range of possible values for $\amub$. The scanned
  parameter space is defined by Eq.\ (\ref{parameterranges}), with
  Eq.\ (\ref{lambda1solution}) in case $\MMA<\MMh/2$. Only points
  passing all constraints of sec.\ \ref{sec:constraints} are shown. Plot (a) would
  remain 
  essentially the same for other choices of $\zl$, and plot (b) would
  change 
  essentially linearly with $\zl$. In plot (b), part of the region
  below $\MMA<20$ GeV is excluded for $\zl=-60$, corresponding to the
  limit in Fig.\ \ref{fig:maximumzl}.}
\end{figure}

Hence the overall bosonic two-loop contributions are essentially
proportional to the value of the coupling $C_{HH^+H^-}$. Likewise, all
the parameters $\tan\beta$, $\lambda_{1,6,7}$ enter the prediction for
$\amu$ essentially via this coupling. This proportionality is shown in
Fig.\ \ref{fig:rho}, which displays the ratio $\rho$, defined via
\begin{align}
  |\amub|&= \rho |C_{HH^+H^-}/\text{GeV}|\,|\zl|\times10^{-15}
\label{rho}
\end{align}
as a function of $\amub$ in a scan of parameter space. The approximate
proportionality clearly emerges, if $\amub$ is larger than around
$0.5\times10^{-10}$. The quantity $\rho$ then only depends on the
heavy Higgs masses, and its value is $\rho\approx6,3,2,1$ (for
$\MMH=\MMHpm=150,200,250,300$ GeV, respectively). In
Fig.\ \ref{fig:rho} we display only positive $\amub$. The sign of
$\amub$ also depends on the triple Higgs coupling (see 
the explicit formula in the appendix). For small $\CBA$ it is thus
determined essentially by $(\tan\beta-1)$. If
$\tan\beta<1$, $\amub$ is positive (for negative $\zl$ and with small
corrections if $\CBA\ne0$). 

Hence we mainly need to discuss the behaviour of the coupling
$C_{HH^+H^-}$. We need to
distinguish two cases:
\begin{itemize}
  \item 2HDM type I, II, X, Y: here $\tan\beta$ and 
    the Yukawa parameters are correlated. Specifically in the most
    interesting case of the type X model, $\tan\beta=-\zl$ and is
    therefore large. As a result, the triple Higgs coupling is
    suppressed, and the overall bosonic contribution $\amub$ is
    negligible.
    \item General aligned 2HDM: in this case $\tan\beta$ is
      independent of $\zl$, and the triple Higgs coupling
      $C_{HH^+H^-}$ can be largest if $\tan\beta={\cal O}(1)$.
\end{itemize}
Focusing now on the second case of the aligned 2HDM, the range of
possible values of $C_{HH^+H^-}$ can already be seen in
Fig.\ \ref{fig:zuscatter} for particular choices of $\MMA=50/80$~GeV, $\MMH=\MMHpm=200$~GeV,
  $\zeta_l=-40$. There, large  $C_{HH^+H^-}$ was important to suppress
the branching ratio of $H\to\tau\tau$ and allow large values for
$\zu$. For $\MMA=80$ GeV all parameters $\lambda_{1,6,7}$ and
$\tan\beta$ have been varied in the full range of
Eq.\ (\ref{parameterranges}), and the maximum  
allowed triple Higgs coupling is around $1000$ GeV. For $\MMA=50$ GeV,
on the other hand, $\lambda_1$ is fixed as explained in
sec.\ \ref{sec:basicconstraints} to suppress the decay $h\to
\MA\MA$. Hence the maximum triple Higgs coupling is smaller, in this
case around 400 GeV.

The results generalize to other values of $\MMA$. The maximum triple
Higgs coupling essentially only depends on whether $\MMA$ is smaller
or larger than $\MMh/2$. In the latter case, the triple Higgs coupling
reaches around $1000$ GeV, in the former case only around $400\ldots
600$ GeV, depending on the heavy Higgs masses $\MMH$, $\MMHpm$.

Figure \ref{fig:multiMamuB} shows the range of possible bosonic
contributions $\amub$ as a function of $\MMA$ for various values of
$\MMH=\MMHpm$. The result is fully understood with the proportionality
(\ref{rho}) and the maximum values for $C_{HH^+H^-}$ just discussed.
We display the result only for a particular value of $\zl$ but we have
checked that the results are exactly linear in $\zl$ as expected. We
have also checked that 
the maximum results do not change significantly if the heavy
Higgs masses are varied independently,
$\MMH\ne\MMHpm$,  or if $\lambda_{6,7}$ are set to
zero.

As a result of the analysis of the individual contributions to $\amub$
and of $C_{HH^+H^-}$ we can now summarize the maximum possible $\amub$ in
the simple approximation formula
\begin{align}
  |\amub|^{\text{max}}&\approx\left\{\begin{array}{c}1\\0.5
  \end{array}
  \right\}\,\rho \,|\zl| \times10^{-12}
\end{align}
where the upper (lower) result holds for $\MMA>\MMh/2$ $(<\MMh/2)$ and
where $\rho=6,3,2,1$ for $\MMH=\MMHpm=150,200,250,300$ GeV, respectively.

%% file: gm22HDM.sec5.tex
\section{Muon \boldmath{$g-2$} in the 2HDM}

In this section we use the previous results on limits on relevant
parameters to discuss in detail the possible values of $\amu$ in the
2HDM, answering the two questions raised in the
introduction. Subsection \ref{sec:amuparameterregions} discusses
$\amu$ as a function of the relevant parameters and characterizes
parameter regions giving particular values for $\amu$; subsection
\ref{sec:maximumamu} provides the  maximum $\amu$ that can be
obtained in the 2HDM overall or for certain parameter values.

Before entering details, we provide here useful approximation formulas
for $\amu$ in the 2HDM, which provide the correct qualitative
behaviour  in the parameter region of interest
with small $\MMA$ and large lepton Yukawa coupling $\zl$. The one-loop
contributions $\amu ^{\thdm, 1}$ are dominated by diagrams with $\MA$
exchange; the fermionic two-loop contributions $\amuf$ are dominated
by diagrams with $\tau$-loop and $\MA$ exchange or top-loop and
$\MA,\MH,\MHpm$ exchange; the 
bosonic two-loop contributions $\amub$ are dominated by diagrams with
$\MH$ exchange and $\MHpm$-loop.
The numerical approximations for these contributions are, using
$\hat{x} _{\cal S} \equiv {M_{\cal S}}/{100\;\text{GeV}}$ and $\MMHpm=\MMH$,  
\begin{subequations}
\label{approximations}
  \begin{align}
\amu ^{\thdm, 1}& \simeq  \Big(\frac{\zl}{100}\Big) ^2\, \Big\{ 
 \frac{-3 - 0.5 \ln( \hat{x} _{\MA} )}{\hat{x} _{\MA} ^2}\Big\} \, \times10 ^{-10} \,,
\\
\amuf{}^{\tau}&\simeq \left(\frac{\zl}{100}\right)^2\, \Big\{
  \frac{8+4\hat{x} _{\MA}^2+2\ln(\hat{x}
    _{\MA})}{\hat{x}_{\MA}^2}\Big\} \,\times10^{-10}\,,
  \\
\amuf{}^{t}&\simeq   
\left(  \frac{-\zl\zu}{100}\right)\, \Big\{
22-14\ln(\hat{x}_{\MA})+32-15\ln(\hat{x}_{\MH})\Big\} \,\times10^{-10}
\,,
\label{amutopapprox}
\\
|\amub|&\simeq \rho|C_{HH^+H^-}/\text{GeV}|\,|\zl|\times10^{-15}\,.
\label{amubapprox}
\end{align}
\end{subequations}
The sign of the $\tau$-loop contribution is positive in our parameter
region; the one-loop contributions are negative but are subdominant
except at very small $\MMA$. The top-loop contribution is positive if
$\zu$ has a sign opposite to $\zl$, which is why we choose $\zl<0$ and
focus on $\zu>0$.
$\amub$ is positive if $\zl<0$ and $\tan\beta<1$ (up to small
corrections if $\amub$ is small); see sec.\ \ref{sec:amub} for further
details on the quantity $\rho$ and the approximation for $\amub$.

For the exact results we refer to the literature. The full two-loop
result has been obtained and documented in Ref.\ \cite{oldpaper}; the
full set of Barr-Zee diagrams has been obtained in
Ref.\ \cite{Ilisie:2015tra}; for earlier results we refer to the
references therein. In our numerical evaluation we use the results of
Ref.\ \cite{oldpaper}.

\subsection{\boldmath{$\amu$} in different parameter regions}
\label{sec:amuparameterregions}

\begin{figure}[ht]
  \begin{center}
  \begin{subfigure}[]{.47\textwidth}
    \includegraphics[width=.99\textwidth]{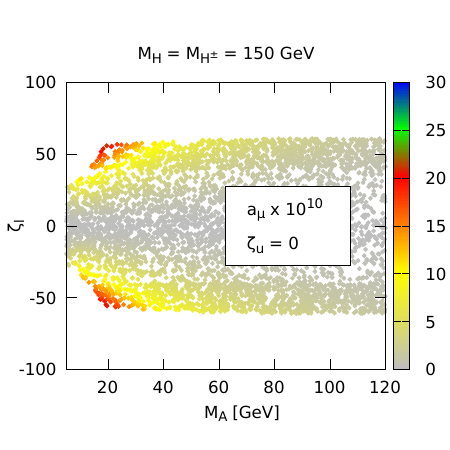}
    \subcaption{\label{fig:amutau150}}
  \end{subfigure}\hfill
  \begin{subfigure}[]{.47\textwidth}
    \includegraphics[width=.99\textwidth]{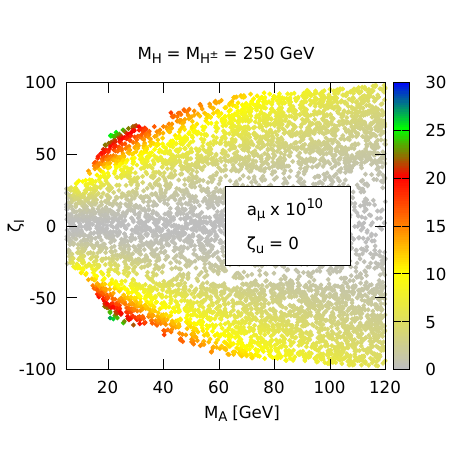}
    \subcaption{\label{fig:amutau250}}
  \end{subfigure}\quad
  \end{center}
  \caption{\label{fig:amutauOL} $\amu$ in the 2HDM (from two-loop fermionic
  and one-loop contributions, and in units of $10^{-10}$), as a function of $\MMA$ and the
  $\tau$-Yukawa parameter $\zl$; the current deviation
  (\ref{deviation}) corresponds to green points. Only points in the allowed region of
  Fig.\ \ref{fig:maximumzl} are shown. The parameter $\zu$ is set to zero,
  corresponding to the case with vanishing top-loop
  contributions and approximately to the type X model case. The parameters $\MMH$,
  $\MMHpm$ are fixed as indicated. Corresponding plots with
  $\MMH,\MMHpm=200,300$ GeV would look very similar, except for the
  slightly different allowed regions.}
\end{figure}

Here we discuss the question raised in the introduction: In which parameter region can the 2HDM accommodate the
      current deviation in $\amu$ (or a future, possibly larger or
      smaller deviation)?

\begin{figure}
  \begin{subfigure}[]{.31\textwidth}
    \includegraphics[scale=.63]{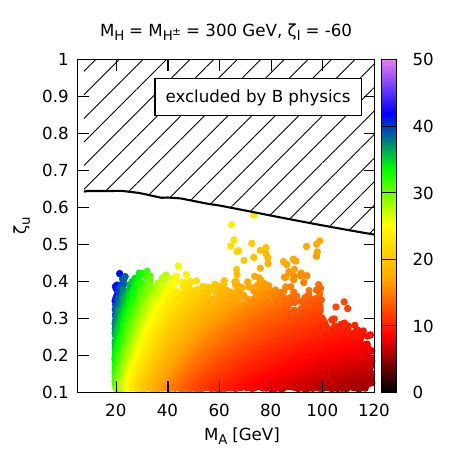}
    \subcaption{\label{fig:amuFOM300C2}}
  \end{subfigure}\quad
  \begin{subfigure}[]{.31\textwidth}
    \includegraphics[scale=.63]{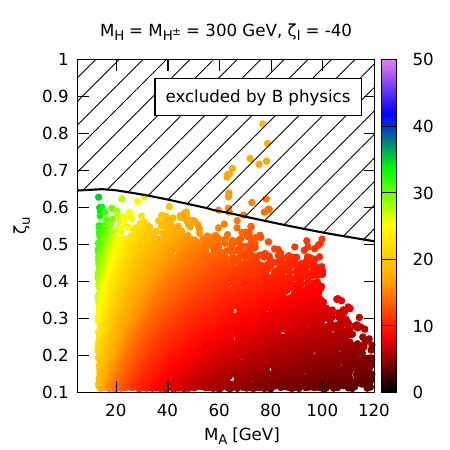} 
    \subcaption{\label{fig:amuFPM300C4}}
  \end{subfigure}\quad
  \begin{subfigure}[]{.31\textwidth}
    \includegraphics[scale=.63]{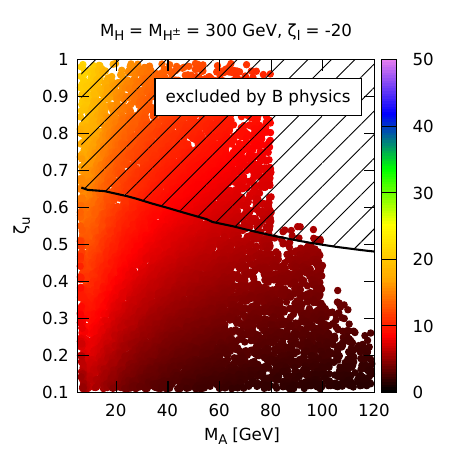} 
    \subcaption{\label{fig:amuFPM300C6}}
  \end{subfigure}
\caption{\label{fig:amuFO} $\amu$ in the 2HDM (from two-loop fermionic
  and one-loop contributions, and in units of $10^{-10}$), as a function of $\MMA$ and the
  top-Yukawa parameter $\zu$; the current deviation
  (\ref{deviation}) corresponds to yellow/green points. Only points
  allowed by the collider constraints of 
  Fig.\ \ref{fig:LHCBmaximumplots} are shown; the B-physics
  constraints are shown as the hatched regions. The parameters  $\zl$ and $\MMH$,
  $\MMHpm$ are fixed as indicated. Corresponding plots with other
  choices of $\MMH$, $\MMHpm$ would look very similar, except for the
  different allowed parameter regions.}
\end{figure}

We begin by listing several remarks which can be obtained from the
results of the previous sections.
\begin{itemize}
  \item All important contributions to $\amu$ are proportional to the
    lepton Yukawa coupling parameter $\zl$ or $\zl^2$ (where e.g.\ in
    the type X model $\zl=-\tan\beta$). Hence $\zl$ must be much
    larger than unity in order to obtain significant $\amu$. From
    section \ref{sec:zetauconstraints} we then obtain that the quark
    Yukawa parameters $\zu$, $\zd$ can be at most of order unity.

    This implies that the bottom loop contribution is negligible, and
    that the type X model is the only of the usual four discrete
    symmetry models with significant $\amu$ (see also
    Ref.\ \cite{ChunPassera}). 
  \item The single most important contribution to $\amu$ is the one
    from the $\tau$-loop, see Eq.\ (\ref{approximations}). It depends on
    $\zl$ and $\MMA$. In the general flavour-aligned model, the
    top-loop contribution can also be significant provided $\zu$ is
    close to its maximum value of order unity.
  \item The masses of the heavy Higgs bosons $\MH$ and $\MHpm$ are
    relatively unimportant for $\amu$. However, they are important for
    the limits on the possible values of $\zl$ and $\zu$. If these
    Higgs bosons have masses around 250 GeV the largest
    $|\zl|$ up to 100 are allowed in most  of the parameter space. For
    even higher masses the limits on $|\zl|$ become slightly stronger and the
    limits on $\zu$ saturate thanks to Z-decay and LHC
    search limits.
  \item The mass splitting between $\MMH$ and $\MMHpm$
    is unimportant. It is strongly restricted by limits from
    electroweak precision observables
    \cite{ChunPassera,Hessenberger:2016atw} and we have 
    checked that its remaining influence on the limits on $\zl$,
    $\zu$ and on the bosonic contributions $\amub$ is
    negligible. Hence we set
    $\MMH=\MMHpm$  in all our numerical examples. 
  \item The Higgs mixing angle $\CBA$ is unimportant for $\amu$. For
    our scenario of interest it is mostly limited by LHC measurements
    of Higgs couplings to leptons, which restrict $|\CBA\zl|$ to be
    smaller than order one. Hence all contributions to $\amu$
    depending on $\CBA$ are strongly suppressed.
  \item The parameters $\lambda_{1,6,7}$ and $\tan\beta$ from the
    Higgs potential appear in $\amu$ essentially only via the triple
    Higgs coupling $C_{HH^+H^-}$, which in turn is maximized for
    $\tan\beta={\cal O}(1)$. In the type X model with large
    $\tan\beta=-\zl$ this strongly suppresses the bosonic
    contributions $\amub$; in the more general aligned model, the
    bosonic diagrams behave as given in Eq.\ (\ref{amubapprox}).

    In the plots of this subsection we do not include the bosonic
    contributions $\amub$ because their parameter dependence is clear
    from this discussion, because their sign can be positive or
    negative, and because their numerical impact is small. 
\end{itemize}
Figures \ref{fig:amutauOL} and \ref{fig:amuFO} show $\amu$ as a
function of the most important parameters $\MMA$, $\zl$ and $\zu$ and
the heavy Higgs masses $\MMH,\MMHpm$.

Fig.\ \ref{fig:amutauOL} focuses on the two most important parameters
$\MMA$ and the lepton Yukawa coupling $\zl$. It shows $\amu$
(including one-loop and fermionic two-loop contributions) as a
function of $\MMA$ and $\zl$. The top-Yukawa parameter $\zu$ is fixed
to $\zu=0$; hence only the $\tau$-loop and the one-loop contributions
are significant. The result also corresponds to
the type X model, in which $\zu$ is negligible. We further fix
$\MMH=\MMHpm=150,250$ GeV and show only parameter points  allowed by the
constraints of sec.\ \ref{sec:zlconstraints}. The results for $\amu$
are not very sensitive to the choice of $\MMH,\MMHpm$, but for
$\MMH=\MMHpm=250$ GeV the allowed parameter space is largest.

Even at the border of the allowed region, a contribution as large as
the deviation (\ref{deviation}) can barely be obtained (see also the
discussions in Refs.\ \cite{Abe:2015oca,ChunKim}). Only in the small
corner with $\MMA\sim20$ GeV and $|\zl|\sim70$,
$\amu$ comes close to explaining Eq.\ (\ref{deviation}). More
generally, the plot reflects the behaviour that $\amu$ is dominated by
the $\tau$-loop which in turn is approximately proportional to $(\zl/\MMA)^2$. A
contribution above approximately $20$ (in units of $10^{-10}$) is possible in the
small region where $|\zl/\MMA|>2\text{GeV}^{-1}$, which is allowed for
around $\MMA\sim20\ldots40$ GeV. Even smaller contributions 
above  $10$ are difficult to obtain. They
require  $|\zl/\MMA|>1\text{GeV}^{-1}$ and are possible for
$\MMA$ up to around $60\ldots80$ GeV.

The impact of the top-loop for $\zu\ne0$ can be seen in
Fig.\ \ref{fig:amuFO}. It shows $\amu$ (including one-loop and fermionic
two-loop contributions) as a function of $\MMA$ and $\zu$. In the
plot, $\zl$ is fixed to exemplary values $\zl=-20,-40,-60$.
Because of the sum of $\tau$- and
top-loops the dependence on $\zl$ is non-linear, and the relative
importance of the top-loop and thus of the parameter $\zu$ is higher
for smaller $\zl$.

We display $\amu$ for all points which pass the collider constraints
discussed in sec.\ \ref{sec:zetauconstraints}, and we display the
constraints from B-physics on the maximum $\zu$ as a line in the
plots. In Fig.\ \ref{fig:amuFO} we do not show all choices of the
heavy Higgs masses $\MMH,\MMHpm$ but fix $\MMH=\MMHpm=300$ GeV. Like
in the previous figure, the values of $\amu$ would be essentially
independent of the heavy 
Higgs masses; the behaviour of the collider and B-physics
constraints can be obtained from Fig.\ \ref{fig:LHCBmaximumplots}.

Nonzero $\zu$ helps in explaining the current $\amu$ deviation
(\ref{deviation}) of around $30$ (in units of $10^{-10}$). The fan-shaped
structure of the plots shows that higher values of the Higgs mass
$\MMA$ can be compensated by larger $\zu$ to obtain the same
$\amu$. For instance, for $\zl=-60$,
contributions to $\amu$ around $30$ can be
obtained up to 
$\MMA\sim40$ GeV. Contributions above $20$ can be obtained up to
$\MMA\sim M_Z$, by taking advantage of the larger allowed
values of $\zu$ in this mass range.

For smaller $\zl=-40$,
contributions above $20$ are possible for $\MMA$ up to around $60$
GeV, and contributions above $10$ are possible up to $\MA\sim M_Z$.
For $\zl=-20$, the contributions to $\amu$ are generally smaller than
$20\times10^{-10}$, but even here nonzero $\zu$ strongly increases
$\amu$.

\subsection{Maximum possible $\amu$ in the 2HDM}
\label{sec:maximumamu}

\begin{figure}
  \begin{subfigure}[]{.5\textwidth}
    \includegraphics[scale=.9]{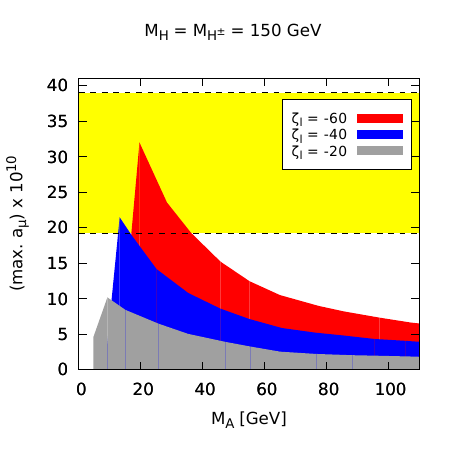}
    \subcaption{\label{fig:amumax150}}
  \end{subfigure}\quad
  \begin{subfigure}[]{.5\textwidth}
    \includegraphics[scale=.9]{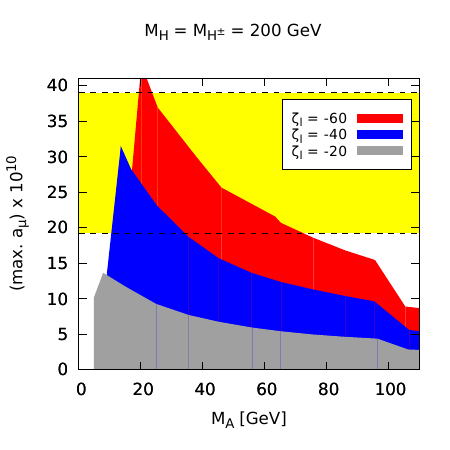} 
    \subcaption{\label{fig:amumax200}}
  \end{subfigure}\vspace{1cm}\\
  \begin{subfigure}[]{.5\textwidth}
    \includegraphics[scale=.9]{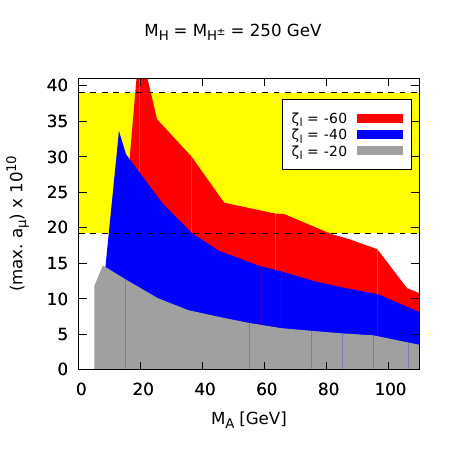}
    \subcaption{\label{fig:amumax250}}
  \end{subfigure}\quad
  \begin{subfigure}[]{.5\textwidth}
    \includegraphics[scale=.9]{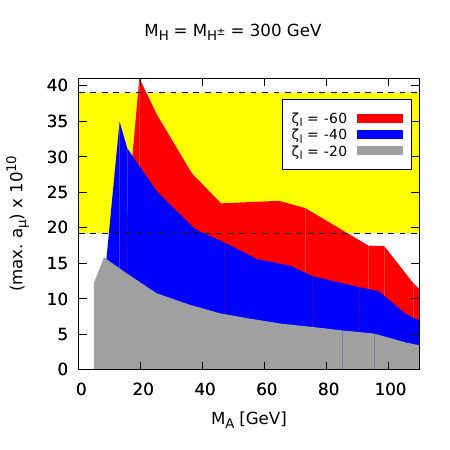} 
    \subcaption{\label{fig:amumax300}}
  \end{subfigure}
\caption{\label{fig:amumaxzl} The maximum 
  $a_{\mu}$ (including one-loop and all two-loop contributions) for
  several fixed values of $\zl$ and $\MMH=\MMHpm$. For each $\MMA$ and
  $\zl$, the maximum $\zu$ is obtained
  from the results of sec.\ \ref{sec:zetauconstraints}. The yellow band
  indicates the current $\amu$ deviation, defined by taking the envelope of
  the $1\sigma$ bands given by Eq.\ (\ref{deviation}).} 
\end{figure}

\begin{figure}
  \begin{subfigure}[]{.5\textwidth}
    \includegraphics[scale=.9]{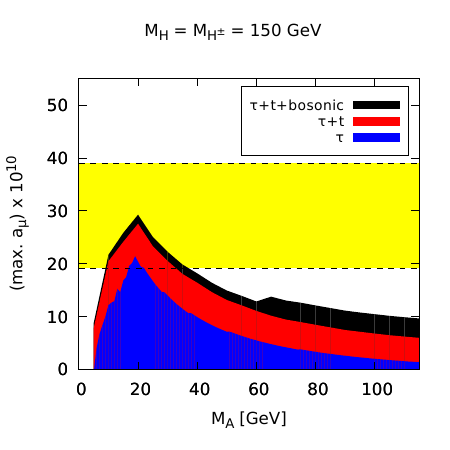}
  \end{subfigure}\quad
  \begin{subfigure}[]{.5\textwidth}
    \includegraphics[scale=.9]{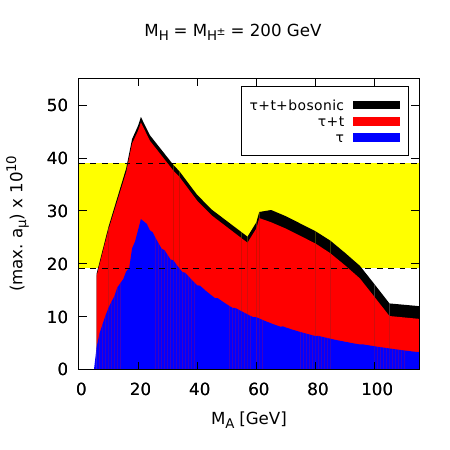} 
  \end{subfigure}\vspace{1cm}\\
  \begin{subfigure}[]{.5\textwidth}
    \includegraphics[scale=.9]{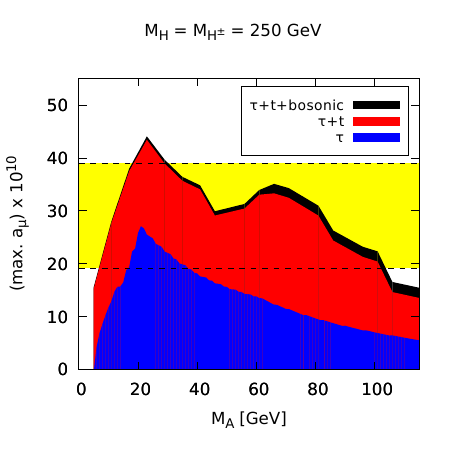}
  \end{subfigure}\quad
  \begin{subfigure}[]{.5\textwidth}
    \includegraphics[scale=.9]{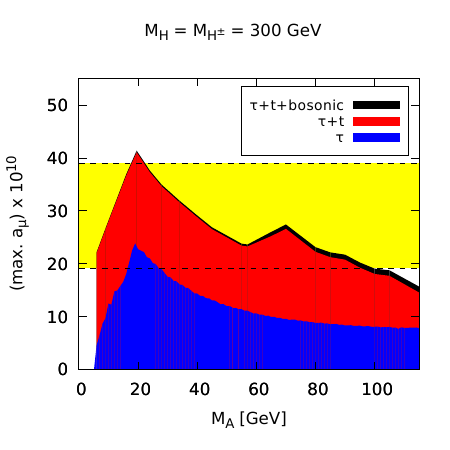} 
  \end{subfigure}
\caption{\label{fig:amumaxoverall} The overall maximum 
  $a_{\mu}$ (including one-loop and all two-loop contributions) as a
  function of $\MMA$, for
  several fixed values of $\MMH=\MMHpm$. For each value of $\MMA$, the
  maximum value of $|\zl|$ is determined as in
  sec.\ \ref{sec:zlconstraints}; then the maximum $\zu$ is obtained
  from the results of sec.\ \ref{sec:zetauconstraints}. The result without top-loop
  and bosonic contributions (which would correspond to the maximum in
  the type X model) is shown in blue; the result without bosonic
  two-loop contributions in red; the total maximum result, including
  the maximum bosonic contributions in black. The yellow band
  indicates the current $\amu$ deviation, defined by taking the envelope of
  the $1\sigma$ bands given by Eq.\ (\ref{deviation}).} 
\end{figure}

Now we discuss the question: What is the overall maximum possible value of $\amu$ that
        can be obtained in the 2HDM? 
Fig.\ \ref{fig:amumaxzl} and \ref{fig:amumaxoverall} show the maximum
possible $\amu$ in the 2HDM, first for fixed choices of the lepton
Yukawa coupling $\zl=-20,-40,-60$, 
then overall.

Fig.\ \ref{fig:amumaxzl} is obtained by maximizing $\zu$ for each parameter point, given 
all constraints discussed in sec.\ \ref{sec:zetauconstraints}.
The plots clearly show the prominent role of $\MMA$ and the lepton
Yukawa coupling $\zl$. The values of the heavy Higgs bosons
$\MMH,\MMHpm$ mainly matter because they influence the maximum
allowed value of $\zu$. Only two cases need to be clearly distinguished: small
$\MMH,\MMHpm=150$ GeV and larger $\MMH,\MMHpm=200,250,300$ GeV,
which all lead to similar results for $\amu$.

For each value of
$\zl$, there is a sharp maximum around $\MMA\sim20$ GeV. At the
maximum, $\amu$ obviously depends on $\zl$, but also on the heavy
Higgs masses $\MMH,\MMHpm$, because their values influence the maximum
allowed value of $\zu$. For $\zl=-60(-40)$ and large $\MMH,\MMHpm$, $\amu$
reaches $40(30)\times10^{-10}$, which is larger than the currently
observed deviation (\ref{deviation}). For $\MMH=\MMHpm=150$ GeV or
$\zl=-20$, the contributions to $\amu$ are smaller.

For values of $\MMA$ lower than at the peaks in Fig.\ \ref{fig:amumaxzl}, the maximum $\amu$
values drop sharply  (the drop is at lower $\MMA$ if
$\zl$ is smaller). The reason is that  for each $\zl$ there is a
minimum allowed value of $\MMA$ mainly because of the 
collider limits discussed in
sec.\ \ref{sec:zlconstraints}. Even if lower values of $\MMA$ were
allowed, $\amu$ would be suppressed by the negative one-loop
contribution. 

For higher values of $\MMA$, $\amu$ is suppressed
by $\MMA$. As can be estimated from the approximation
(\ref{approximations}), the suppression is weaker than
$1/\MMA^2$. Further the suppression is modulated by the maximum 
possible value of $\zu$. In particular, above $\MMA>\MMh/2$, higher
values of $\zu$ are allowed, and the maximum $\amu$ drops more slowly
with $\MMA$.

In summary, the deviation (\ref{deviation}) can be explained at the
$1\sigma$ level for $\MMA=20\ldots40$ GeV and for $\zl=-40$ and high
$\MMH,\MMHpm$ or $\zl=-60$ independently of $\MMH,\MMHpm$. It can
further be explained for $\MMA=20\ldots80$ GeV for $\zl=-60$ if  
$\MMH,\MMHpm$ are high.

The overall maximum $\amu$ in the flavour-aligned 2HDM can be seen in
Fig.\ \ref{fig:amumaxoverall} for several choices of $\MMH,\MMHpm$.
The figure is obtained by maximizing first $\zl$ (i.e.\ the
$\tau$-loop contribution), then $\zu$ (i.e.\ the top-loop
contribution), and finally the bosonic two-loop contribution for each
parameter point. All constraints discussed in secs.\ \ref{sec:zlconstraints},
\ref{sec:zetauconstraints} are employed.

The plots display 
not only the final total result for $\amu$ including
all one- and two-loop contributions. They also display the results
of the $\tau$-loop (plus one-loop)
contribution alone, and the results including the top-loop but
excluding the bosonic two-loop contributions. In this way the plots
allow to read off the results corresponding to the 2HDM type X, and to
read off the influence of the bosonic two-loop corrections.

Starting the discussion with the type X model result (blue), the plots
confirm that the type X model can barely explain the current deviation
(\ref{deviation}). The largest values that can be obtained are around
$27\times10^{-10}$ at $\MMA=20$ GeV for $\MMH,\MMHpm=200\ldots 250$
GeV. For higher or lower values of $\MMA$ the maximum type X
contributions drop quickly, and values above $20\times10^{-10}$ can
only be obtained between $\MMA=20\ldots40$ GeV.

Hence going beyond the type X model and allowing general Yukawa
couplings significantly widens the parameter space which can lead to
significant contributions to $\amu$. Both the top-loop 
and the bosonic two-loop contributions can significantly increase
$\amu$. Thanks to the behaviour discussed in sec.\ \ref{sec:amub} and
expressed in Eqs.\ (\ref{approximations}) both of these
contributions are not significantly suppressed by heavier $\MMA$. On
the contrary, for heavier $\MMA$, larger $\zu$ and larger triple Higgs
couplings $C_{HH^+H^-}$ are allowed, and the loop functions are not
strongly suppressed by heavy $\MMA$.

Thus, in the general (flavour-)aligned 2HDM one can obtain even
$\amu>45\times10^{-10}$ if $\MMA\sim20$ GeV and if $\MMH,\MMHpm$ are
in the range 200\ldots250 GeV. Hence the 2HDM could even accommodate a
larger deviation than (\ref{deviation}), which might be established at
forthcoming $\amu$ measurements.
Thanks to the large possible values of the top Yukawa parameter $\zu$,
the current deviation can be explained at the $1\sigma$ level in all
the range $\MMA=20\ldots100$ GeV.

\section{Summary and conclusions}
\label{sec:conclusions}

The 2HDM is a potential source of significant contributions to the
anomalous magnetic moment of the muon $\amu$, and it could explain the
current deviation (\ref{deviation}). Here we have provided a
comprehensive analysis of the relevant parameter space and of possible
flavour-aligned 2HDM contributions to $\amu$. Our analysis was kept general,
anticipating that future $\amu$ measurements might further increase or
decrease the deviation (\ref{deviation}).

The relevant parameter space is characterized by light pseudoscalar
Higgs with mass $\MMA<100 $ GeV and large Yukawa couplings to
leptons. Among the usual 2HDM models with discrete symmetries this is
only possible in the lepton-specific type X model. In the type X
model, large lepton Yukawa couplings imply negligible quark Yukawa
couplings to the $\MA$ boson. We considered the
more general flavour-aligned model, which contains type X as a special
case but in which simultaneously
significant Yukawa couplings to quarks are possible.

We first investigated the allowed values of the Yukawa coupling
parameters $\zl$ and $\zeta_{u,d}$ (which would be given by
$-\tan\beta$ and $1/\tan\beta$ in the type X model). An extensive
summary of the results is provided at the beginning of
sec.\ \ref{sec:amuparameterregions}. In short, the lepton Yukawa
coupling $|\zl|$ can take values up to $40\ldots100$, depending on the
values of all Higgs masses. For very light $\MMA<20$ GeV, very severe
limits from LEP data reduce the maximum $|\zl|$ and thus the maximum
$\amu$. For large lepton Yukawa coupling, both quark Yukawa couplings
$\zeta_{u,d}$ can be ${\cal O}(0.5)$ at most because of B-physics data
and LHC-Higgs searches. While $\zd$ has
negligible influence on $\amu$, in
particular the upper limit on the top Yukawa coupling $\zu$ is
critical for $\amu$. Interestingly, for $\MMA>\MMh/2$ GeV, slightly larger
values of $\zu$ are allowed  thanks to an interplay between the triple
Higgs couplings and the Yukawa coupling.

As an intermediate result and an update of the results of
Ref.\ \cite{oldpaper} on the full two-loop calculation of $\amu$ in
the 2HDM, we evaluated the maximum contributions $\amub$ from
bosonic two-loop diagrams. Going beyond the type X model can
also increase $\amub$. The maximum is mainly determined by the
maximum  triple Higgs coupling, which is obtained if
$\tan\beta\ll|\zl|$. It
reaches $3\times10^{-10}$ if $\zl$ is also maximized and if $\MMA$ is
around 100 GeV and the heavy 
Higgs masses $\MMH,\MMHpm$ are not much higher.

Figures
\ref{fig:amutauOL},\ref{fig:amuFO},\ref{fig:amumaxzl},\ref{fig:amumaxoverall} 
answer the questions how $\amu$ depends on the 2HDM parameters, and 
what is the maximum $\amu$ that can be obtained in the 2HDM.
The overall maximum is above $45\times10^{-10}$, and it can be
obtained for $\MMA\sim20$ GeV. More generally contributions
significantly above the current deviation (\ref{deviation}) can be
obtained for $\MMA$ up to $40$ GeV. Thanks to the large
allowed top Yukawa coupling, the current deviation (\ref{deviation})
can be explained at the $1\sigma$ level for $\MMA$ up to 100 GeV.  Even
if the lepton Yukawa coupling  is not maximized but fixed at only
$\zl=-40$, a $1\sigma$ explanation is possible up to $\MMA=40$
GeV. The heavy 
Higgs masses $\MMH$ and $\MMHpm$ are not very critical; the maximum
$\amu$ is obtained if they are in the range 200\ldots300 GeV; for
lower or higher masses the limits on the Yukawa couplings become
stronger, and significantly higher masses are disfavoured by
triviality constraints \cite{ChunPassera,Abe:2015oca}.

For the type X model, the maximum contributions are significantly
smaller, only slightly above $25\times10^{-10}$. A $1\sigma$
explanation of the current deviation is only possible in the small
range of $\MMA$ between 20 and 40 GeV, and even a potential future
deviation of only $10\times10^{-10}$ can be explained only for
$\MMA<80$ GeV.

In view of these results it is of high interest to test this parameter
space more fully at the LHC. In view of the significant couplings of
the low-mass $\MA$ boson to $\tau$ leptons and top quarks, it is
promising to derive more stringent upper limits on these couplings,
particularly on the product $|\zl\zu|$. Such more stringent limits
will have immediate impact on the possible values of $\amu$ in the
2HDM. At the same time, the future $\amu$ measurements have a high
potential to constrain the 2HDM parameter space. In particular the
type X model might be excluded by a confirmation of a large $\amu$
deviation, and in the more general model, lower limits on the top
Yukawa coupling and upper limits on $\MMA$ might be
derived\footnote{Here we comment on
  Ref.~\cite{Keus:2017ioh}, which appeared shortly after the present
  paper, and which claims that large $a_\mu$ is possible for large
  $\MMA$ in case of CP violation. We point out that the large
  $a_\mu$ does not result from CP violation but from (extremely) large considered values of $\tb$($\cot\beta$). However these large
  $\tb$($\cot\beta$) values are excluded by either LHC or B-Physics results
  and therefore not considered in the present paper.}.

%% file: gm22HDM.app.tex
\begin{appendix}
  \section{Explicit results for triple Higgs couplings}
Here we provide the explicit results for the triple Higgs couplings
which are required for our analysis. The triple Higgs couplings of the
heavy Higgs $\MH$ to either $\MA\MA$ or $\MHpm\MH^{\mp}$ are correlated as
\begin{align}
  \label{eq:CHHpHm}
   C_{H H^+ H^-} =& C_{HAA} - 2\left(\frac{\MMHpm ^2 - \MMA ^2}{v}\right)\CBA\,,
\end{align}
and $C_{HAA}$ is given by
\begin{align}
  \label{eq:CHAA}
  C_{HAA} =& \lambda_1 v \Big( \SBA\frac{1-\tb^2}{\tb^3} - \CBA\frac{2}{\tb^{2}} \Big)
  + \SBA\frac{\MMh ^2}{v}\frac{\tb^2-1}{\tb^3}
  + \CBA\Big(\frac{\MMh ^2}{v}\frac{2+\tb^2}{\tb ^{2}} - 2\frac{\MMA ^2}{v}\Big) \nonumber\\
  +& \CBA\frac{\MMh ^2-\MMH ^2}{v}\Big(\frac{2}{\tb^2}- 3 + \CBA\SBA\frac{1-6\tb^2+\tb^4}{\tb^3} + 4\CBA^2\frac{\tb^2-1}{\tb^2}\Big) \nonumber\\
  +& \lambda_6 v \Big(\SBA\frac{2-\tb^2}{\tb^2} - \CBA\frac{3}{\tb}\Big) + \lambda_7 v (-\SBA + \CBA\tb)\,.
\end{align}
The triple Higgs coupling relevant for the potential SM-like Higgs
decay $\Mh\to\MA\MA$ is given by
\begin{align}
  \label{eq:hAA}
  C_{hAA} =& \lambda_1 v \Big(\CBA\frac{\tb^2 - 1}{\tb^3} - \SBA\frac{2}{\tb^2}\Big)
  + \CBA\frac{\MMh ^2}{v}\frac{1-\tb^2}{\tb^3}
  + \SBA\Big(\frac{\MMh ^2}{v}\frac{2+\tb^2}{\tb^2} -2\frac{\MMA ^2}{v} \Big) \nonumber\\
  +& \frac{\MMh ^2 - \MMH ^2}{v}\Big( \CBA\frac{\tb^2-1}{\tb^3} + \CBA\SBA^2\frac{1-6\tb^2+\tb^4}{\tb^3} - 2\SBA + 4\SBA\CBA ^2\frac{\tb^2-1}{\tb^2} \Big) \nonumber\\
  +& \lambda_6 v \Big(\CBA\frac{\tb^2 - 2}{\tb^2} - \SBA\frac{3}{\tb}\Big) + \lambda _7 v (\CBA + \SBA\tb)\,.
\end{align}

\end{appendix}